\documentclass[useAMS,usenatbib]{mn2e}

\usepackage{savesym}
\usepackage{amsmath}
\savesymbol{iint}
\usepackage{txfonts}
\restoresymbol{TXF}{iint}

\usepackage{mathabx}
\usepackage{graphicx}

\usepackage{fixltx2e}

\title[Geothermal heating on synchronously rotating planets]{Geothermal heating enhances atmospheric asymmetries on synchronously rotating planets}

\author[Haqq-Misra and Kopparapu]{Jacob Haqq-Misra$^{1,2}$\thanks{E-mail: 
jacob@bmsis.org} and Ravi Kumar Kopparapu$^{1,2,3,4,5}$ \\
$^{1}$Blue Marble Space Institute of Science, 1200 Westlake Ave N Suite 1006, Seattle, WA 98109, USA\\
$^{2}$Virtual Planetary Laboratory, PO Box 351580, Seattle, WA 98195, USA\\
$^{3}$Department of Geosciences, Pennsylvania State University, 443 Deike Building, University Park, PA 16802 USA\\
$^{4}$Center for Exoplanets \& Habitable Worlds, Pennsylvania State University, 525 Davey Lab, University Park, PA 16802 USA\\
$^{5}$Penn State Astrobiology Research Center, Pennsylvania State University, 2217 Earth and Engineering Sciences Building, University Park, PA 16802 USA}

\begin{document}

\date{This version 2014 July 4}

\pagerange{\pageref{firstpage}--\pageref{lastpage}} \pubyear{2014}

\maketitle

\label{firstpage}

\begin{abstract}
Earth-like planets within the liquid water habitable zone of M type stars may evolve into synchronous rotators. 
On these planets, the sub-stellar hemisphere experiences perpetual daylight while the opposing anti-stellar 
hemisphere experiences perpetual darkness. Because the night-side hemisphere has no direct source of 
energy, the air over this side of the planet is prone to freeze out and deposit on the surface, 
which could result in atmospheric collapse. However, general circulation models (GCMs) 
have shown that atmospheric dynamics can counteract this problem and provide sufficient energy transport to the 
anti-stellar side.

Here we use an idealized GCM to consider the impact of geothermal heating on the habitability of synchronously 
rotating planets. Geothermal heating may be expected due to tidal interactions with the host star, and the effects 
of geothermal heating provide additional habitable surface area and may help to induce melting of ice 
on the anti-stellar hemisphere. We also explore the persistence of atmospheric asymmetries between the northern 
and southern hemispheres, and we find 
that the direction of the meridional circulation (for rapidly rotating planets) or the direction of zonal 
wind (for slowly rotating planets) reverses on either side of the sub-stellar point. We show that 
the zonal circulation approaches a theoretical state similar to a Walker circulation only for 
slowly rotating planets, while rapidly rotating planets show a zonal circulation with the opposite 
direction. We find that a cross-polar circulation is present in all cases and provides 
an additional mechanism of mass and energy transport from the sub-stellar to anti-stellar point. Characterization of the 
atmospheres of synchronously rotating planets should include consideration of hemispheric 
differences in meridional circulation and examination of transport due to cross-polar flow.
\end{abstract}


\section{INTRODUCTION}

Low-mass M stars are among the most numerous stellar type in the galaxy and also 
among the most long lived. This would make them attractive candidates in the search 
for life beyond Earth were it not for the possibility of atmospheric loss due to 
synchronous rotation. The ongoing discovery of extrasolar planets around 
low-mass stars \citep{basri2005,udry2007,vonbloh2008,bonfils2013,angladaescude2013}
has renewed discussion of the habitability of planets orbiting M type stars \citep{tarter2007}
and provides additional motivation for understanding the atmospheric dynamics of such planets.  

The observation of extrasolar planets around stars of all types has increased interest 
in characterizing planetary atmospheres through climate modeling. These 
include general circulation model (GCM) simulations of 
atmospheric circulation on large, Jupiter-sized planets in close 
orbit to their parent star known 
as ``hot Jupiters" \citep{cooper2005,fortney2006,showman2008,menou2009,showman2011}, 
as well as consideration of smaller, Neptune-sized gas giants 
and Earth-sized terrestrial planets \citep{williams1982,heng2011b}. 
The discovery of planets orbiting the M type star Gliese 581 
also prompted an assessment of the habitability of these new worlds that include radiative climate calculations 
of Gliese 581c \citep{hu2011}. (Note that radiative and GCM studies also were 
conducted for the suspected planets Gliese 581d \citep{vonparis2010,hu2011} and 
Gliese 581g \citep{heng2011,pierrehumbert2011}, which have now been determined 
to be artifacts due to stellar activity \citep{robertson2014}.) These modeling studies 
help to identify prominent circulation features in the atmospheres of extrasolar 
planets and characterize the types of atmospheres that could exist in orbit around 
distant stars

One way to characterize planetary habitability is the 
liquid water habitable zone \citep{kasting1993,kopparapu2013,kopparapu2014}, which describes the region 
around a star where a planet could sustain surface liquid water. This habitable zone 
defines a region where the negative climate feedback of the carbonate-silicate cycle \citep{walker1981} can keep
a planet above freezing. It is  bounded on the inner edge by the loss of water through initiation of a runaway greenhouse effect \citep{kasting1988} 
and on the outer edge by the condensation of carbon dioxide \citep{kasting1991}. Although this habitable 
zone cannot be generalized to all extrasolar planets, it can be used to characterize the habitability of 
any terrestrial planet with water and a carbonate-silicate cycle. 
(Some authors have extended consideration of habitability 
to scenarios such as a dry planet with limited water \citep{abe2011,zsom2013} or a planet rich 
in atmospheric hydrogen \citep{pierrehumbert2011b}.) For low-mass M type stars, this liquid 
water habitable zone falls within the tidal locking orbital distance of its parent 
star, so any potentially habitable planets around these stars 
(according to this definition of habitability) are at risk of atmospheric loss and may in fact be uninhabitable. 

The problem of atmospheric stability for planets orbiting low-mass stars has been investigated
by \citet{joshi1997}, \citet{joshi2003}, \citet{merlis2010}, \citet{edson2011}, \citet{pierrehumbert2011}, 
\citet{heng2011}, \citet{selsis2011}, \citet{edson2012} 
\citet{yang2013}, \citet{leconte2013}, and \cite{yang2014}
using GCMs to assess the degree to which atmospheric circulations can maintain 
planetary habitability. These simulations show that a range of planetary
atmospheres can remain stable under synchronous rotation, either by a sufficiently
dense carbon dioxide atmosphere \citep{joshi1997,joshi2003,selsis2011,edson2012} or
in regimes where atmospheric dynamics provides sufficient heat transport
to the night side \citep{merlis2010,edson2011,pierrehumbert2011,heng2011}.
These GCM studies demonstrate that extrasolar planets orbiting low-mass
stars can indeed provide habitable conditions at locations on the
surface, even with the problem of synchronous rotation.

In this study we use an idealized GCM to investigate the influence of geothermal 
heating on atmospheric dynamics. We present a set of experiments that include fast 
and slow rotators to show that geothermal heating amplifies asymmetries in atmospheric 
circulation between the northern and southern hemispheres, 
which can cause transient regions of warming near the anti-stellar point. 
We show that the dynamical structure of the atmosphere changes on either side of the
sub-stellar point, and we find that a cross-polar circulation provides transport 
from the day to night side in all experiments. These results provide qualitative 
descriptions of the expected meridional, zonal, and cross-polar circulation patterns 
that can help to guide further GCM studies of synchronously rotating planets.

\section{SYNCHRONOUS ORBITS AROUND LOW-MASS STARS}

A planet can fall into synchronous rotation around its parent star when tidal dissipation 
slows down the planet's rotation rate. Not all planets orbiting at this tidal locking distance from 
the parent star will necessarily become synchronous rotators, such as those with eccentric orbits 
or with close neighbors so that they fall into spin-orbit resonances. For example, the planet
Mercury is within the tidal locking distance of our sun, but it has fallen
into a 3:2 spin-orbit resonance that remains stable because of 
its nonzero eccentricity. Planets with more circular
orbits, however, may evolve into synchronous rotators so that a single
side is fixed at perpetual daylight, while the opposing side experiences
perpetual night. 

The tidal migration history of any particular planet depends on 
factors such as the initial rate of rotation, eccentricity, and semi-major axis  
\citep{jackson2008,jackson2008b,jackson2008c,barnes2009}, which complicates 
efforts to determine a particular orbital distance at which a planet should be 
expected to fall into a synchronously rotating state. No single analytic 
equation can describe the tidal locking radius for all planetary systems, and 
empirical formulae may be limited in their predictive power. In this section
we use such an empirical equation to demonstrate limits on stellar mass 
and rotation rate for these systems; however, we admit that this approach 
is simplistic and neglects many key factors that contribute to the actual tidal 
history of a planetary system.

The tidal locking distance $r_T$ can be described empirically 
\citep{peale1977,kasting1993,dobrovolskis2009,edson2011} according to the formula

\begin{equation}
r_{T}=0.024\left(\frac{P_{0}t}{\mathbb{Q}}\right)^{1/6}M^{1/3},
\label{eq:tidallocking}\end{equation}
where $M$ is the mass of the parent star, $P_0$ is the original rotation period of the planet, 
$t$ is the time period since planetary formation, and $\mathbb{Q}$ is a parameter describing 
the inverse of the specific dissipation function for the entire planet. We restrict our 
focus in this study to synchronously rotating Earth-like planets, which allows us to specify 
the parameters in Eq. (\ref{eq:tidallocking}). We assume early Earth rotated more rapidly 
than today so that $P_0 = 13.5\text{ hr}$ \citep{walker1986,kasting1993,williams2000,edson2011}, 
and we likewise use the 4.5 billion year formation period of Earth as the value of $t$. The 
parameter $\mathbb{Q}$ has a relatively low value on Earth today ($\sim$13) because most tidal 
dissipation occurs from gravitational interactions with the sun and Moon along oceanic 
shorelines \citep{kasting1993}. However, this present-day value may be unusually 
large \citep{kasting1993,edson2011}, and
different continental configurations on early Earth---such 
as a larger ocean \citep{rosing2010}--could have reduced the tidal dissipation rate. We will 
therefore use a value of $\mathbb{Q} = 100$ as characteristic of early Earth 
\citep{burns1986,kasting1993,edson2011} to describe tidal forces on a synchronously rotating 
Earth-like planet. The value of the tidal locking distance $r_T$ in Eq. (\ref{eq:tidallocking}) 
is only weakly sensitive to changes in the parameters $P_0$, $t$, and $\mathbb{Q}$, so 
the primary contributor is the mass $M$ of stellar types that could induce synchronous rotation for orbiting 
Earth-like planets.

We consider an Earth-like planet to be a terrestrial planet that receives the same incident flux 
of stellar radiation as Earth. The stellar flux $F$ reaching a planet at an orbital radius $r$ is 
related to the luminosity of its star $L$ according to $F = L / (4\pi r^2)$. We can examine planets 
in orbit around different stellar types by defining the Earth equivalent distance of a planet $r_{EED}$ 
as the orbital distance at which a planet receives an equal magnitude of radiation from its parent star 
as Earth does from the sun \citep{edson2011}. At the Earth equivalent distance, the flux equals that on 
Earth, so we can write 

\begin{equation}
\frac{L}{L_{\Sun}}=\left(\frac{r_{EED}}{r_{\Earth}}\right)^{2},
\label{eq:EED}\end{equation}
where $L_{\Sun}$ is the luminosity of the sun, and $r_{\Earth}$ is the orbital radius of Earth. 
If we use the general approximation that stellar 
luminosity for low-mass stars is proportional to mass to the fourth power $L \sim M^4$, 
then we can combine Eqs. (\ref{eq:tidallocking}) and (\ref{eq:EED}) by setting $r_T \le r_{EED}$
and solving for stellar mass. Letting $M_{\Sun}$ be the mass of the sun, we find that 

\begin{equation}
\frac{M}{M_{\Sun}}\le\left[0.024\left(\frac{P_{0}t}{\mathbb{Q}}\right)^{1/6}\frac{M_{\Sun}^{1/3}}{r_{\Earth}}\right]^{3/5}.
\label{eq:starmass}\end{equation}
Substituting in values for the parameters in Eq. (\ref{eq:starmass}) shows that the Earth equivalent distance $r_{EED}$ and the 
tidal locking distance $r_T$ intersect at $M/M_{\Sun} \approx 0.6$; planets in orbit around stars less 
massive than this will be in synchronous rotation at the Earth equivalent distance. This means that F, G, and K type 
stars are unlikely to harbor Earth-like planets in synchronous rotation, but smaller M type stars are 
ideal candidates to find such planets. 

We use this mass limit to calculate the maximum rotation rate for a synchronously rotating Earth-like planet. Kepler's third 
law gives the relation $(r_{EED}/r_{\Earth})^3 (M_{\Sun}/M) = (P_{EED}/P_{\Earth})^2$, where $P_{EED}$ and 
$P_{\Earth}$ are the orbital period at the Earth equivalent distance and at 
Earth's present distance, respectively. This corresponds to an upper limit on orbital period (equivalently,
rotation rate) of $P_{EED} \approx 230\text{ days}$ for synchronously rotating Earth-like planets around M type stars (c.f. 
\citet{edson2011}). Planets rotating faster than this limit may still be prone to synchronous rotation, while 
planets rotating slower than this limit will be beyond the tidal locking distance and outside the 
liquid water habitable zone \citep{kopparapu2013,kopparapu2014}. In this study, we consider 
synchronously rotating planets with both 1-day ("fast") and 230-day ("slow") rotation rates to show the similarities 
and differences that arise in these two dynamical regimes.

\section{GENERAL CIRCULATION MODEL}

We use a general circulation model (GCM) to simulate the atmospheres of synchronously 
rotating Earth-like planets and investigate the influence of geothermal heating. This GCM 
uses a hydrostatic spectral dynamical core (with T42 resolution), two-stream gray radiative transfer, 
a diffusive boundary layer scheme, and a surface slab ocean \citep{frierson2006,haqqmisra2011}. 
We set obliquity to zero and adjust the incoming solar radiation profile so the sub-stellar point 
is fixed on the intersection of the equator (0$^{\circ}$ latitude) and international 
date line (180$^{\circ}$ longitude). We also include a geothermal heat flux in the GCM 
by adding a constant term to the surface temperature tendency equation. 

This model is still highly idealized, as the model atmosphere is 
transparent to incoming stellar radiation and absorbs only infrared radiation emitted upward from the surface. 
Water vapor is assumed to be radiatively neutral, so this GCM does not include any water vapor feedback, and 
the model experiments also do not include any topography or cloud formation. Although some of these processes 
will likely be significant in determining specific features of climate, these idealizations help to highlight 
large-scale dynamical patterns and make the model an ideal tool for investigating the prominent circulation 
patterns that arise on synchronously rotating planets. 

We consider both dry and moist regimes for our GCM experiments. We force dry GCM conditions by 
setting the saturation vapor pressure equal to zero and using dry adiabatic adjustment. The moist
regime includes latent heat release through large-scale condensation, and a simplified shallow penetrative 
adjustment also known as Betts-Miller adjustment \citep{frierson2007}. For non-synchronously rotating Earth-like 
planets, the presence of moisture in this GCM acts to increase static stability and reduce 
circulation strength \citep{haqqmisra2011}. These idealized dry and moist configurations are designed 
to emphasize the key differences in atmospheric dynamics that arise from the presence or absence 
of water vapor.

We conduct a series of GCM experiments using fast (1-day) and slow (230-day) rotation rates with 
25 vertical levels and physical properties identical to those used by \citet{haqqmisra2011}. 
We initialize dry and moist control states at each rotation rate with a constant temperature 
at all grid points and run the GCM for a 3000 day spin-up period to reach a statistically steady state. 
Geothermal heating is then enabled (or kept disabled), and the model is allowed to continue adjusting for 
another 500 days. We then use the subsequent 1000 days of model output in the analysis that follows. 
Running this GCM using a Linux workstation (6-core Intel E5-2620 Xeon 2.0 GHz) requires approximately 
8 hours in real-time to calculate 1000 model days.

Table \ref{table:experiments} lists the eight GCM experiments we consider in this study. We use 
the designators ``fast'' and ``slow'' in reference to 1-day and 230-day rotation periods, respectively, and  
we likewise indicate ``dry'' and ``moist'' cases accordingly. We use ``control'' to refer to experiments 
without any geothermal flux, and ``geothermal'' indicates the presence of a 2.0 W m$^{-2}$ geothermal 
heat flux across the entire surface. This set of eight GCM experiments provide an overview of expected 
large-scale dynamical structures and the resulting perturbations that arise from geothermal heating.

\begin{table*}
    \centering
    \begin{minipage}{140mm}
    \caption{A list of GCM experiments conducted for this study.}
    \label{table:experiments}
    \begin{tabular}{|l|l|l|l|l|}
        \hline
        Experiment Name       & Rotation Rate & Convective Adjustment & Hydrologic Cycle         & Geothermal Heating \\ \hline
        Fast Dry Control      & 1 day         & dry adiabatic         & none                     & -                  \\ 
        Fast Dry Geothermal   & 1 day         & dry adiabatic         & none                     & 2.0 W m$^{-2}$        \\ 
        Fast Moist Control    & 1 day         & shallow Betts-Miller  & large-scale condensation & -                  \\ 
        Fast Moist Geothermal & 1 day         & shallow Betts-Miller  & large-scale condensation & 2.0 W m$^{-2}$        \\ 
        Slow Dry Control      & 230 days      & dry adiabatic         & none                     & -                  \\ 
        Slow Dry Geothermal   & 230 days      & dry adiabatic         & none                     & 2.0 W m$^{-2}$        \\ 
        Slow Moist Control    & 230 days      & shallow Betts-Miller  & large-scale condensation & -                  \\ 
        Slow Moist Geothermal & 230 days      & shallow Betts-Miller  & large-scale condensation & 2.0 W m$^{-2}$        \\
        \hline
    \end{tabular}
    \end{minipage}
\end{table*}

\section{GEOTHERMAL HEATING}

Terrestrial planets in synchronous rotation around low-mass stars experience tidal heating from gravitational 
interactions with the host star \citep{jackson2008,jackson2008b,jackson2008c,barnes2009,heller2013,barnes2013}. Within the Solar System, Io 
exhibits strong tidal interactions as it orbits Jupiter, which drives constant volcanism across the moon's surface. 
This form of tidal heating also should occur for extrasolar planets in synchronous rotation 
\citep{jackson2008,barnes2009,barnes2013} and even for some exomoons \citep{heller2013}. The magnitude of tidal 
heating for any particular planet depends on its orbital geometry and tidal migration history 
\citep{jackson2008b,jackson2008c}, but some degree of tidal heating should be expected to arise on most  
synchronously rotating planets. 

We consider the effects of tidal heating in the GCM as a geothermal heat flux applied uniformly across all 
surface grid points. Actual profiles of tidal heating on synchronously rotating planets likely will show 
latitudinal and longitudinal variations, but we choose this idealized representation as a first-order 
estimation of the influence of a geothermal heat flux. We choose a value of 2.0 W m$^{-2}$ for the 
geothermal heat flux in our model, which approximately corresponds to the tidal heat flux on Io \citep{heller2013}. 
We do not represent any additional internal heat fluxes, such as from radiogenic decay, so these calculations 
can be considered as an upper limit on the influence of tidal heating on surface temperature 
and atmospheric circulation. 

\begin{figure*}
  \includegraphics[width=40pc]{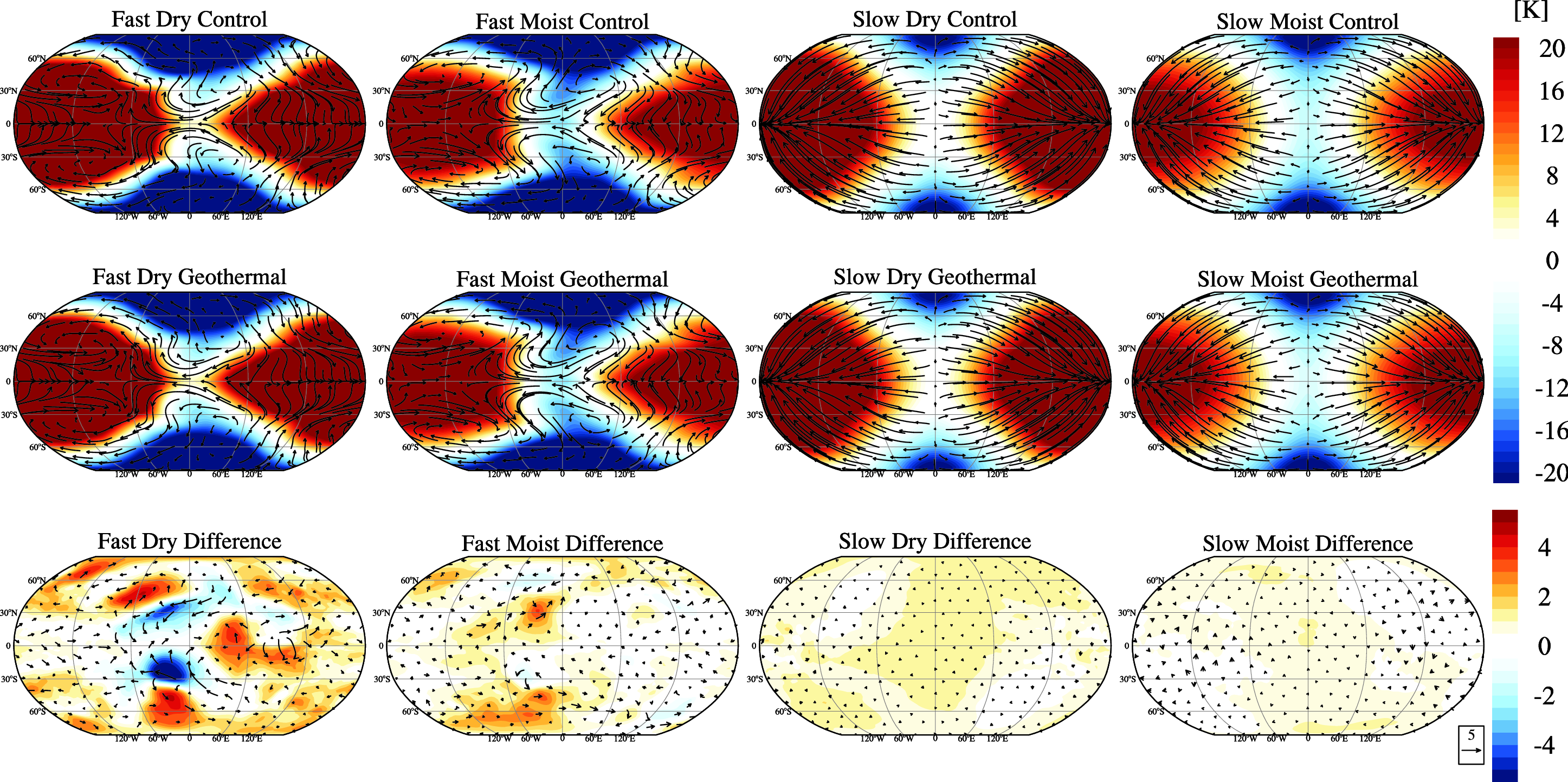}
  \caption{Time average of surface temperature deviation from the freezing point of water (shading) 
and horizontal wind $\overline{u},\overline{v}$ 
(vectors) for a synchronously rotating planet for the (first column)
fast dry, (second column) fast moist, (third column) slow dry, and (fourth column) slow moist 
experiments. The top row shows control cases with no geothermal flux, the middle row shows 
geothermal cases with a 2.0 W m$^{-2}$ geothermal flux, and the bottom row shows the difference 
between geothermal and control cases. The sub-stellar point is centered on the equator along the 
international date line, and each panel is centered on the anti-stellar point (0$^{\circ}$ latitude 
and longitude).}
  \label{fig:surftemp}
\end{figure*}

Our GCM experiments include control cases (no geothermal heat flux) and geothermal cases (2.0 W m$^{-2}$ 
geothermal heat flux) run under dry and moist conditions at both fast and slow rotation rates 
(Table \ref{table:experiments}). Fig. \ref{fig:surftemp} shows the time average of surface temperature
and horizontal wind $\overline{u},\overline{v}$ for the 
full set of experiments. Note that here and elsewhere we show surface temperature 
as the deviation from the freezing point of water. The top row of 
Fig. \ref{fig:surftemp} shows the control cases, the middle row shows the geothermal cases, and the 
bottom row shows the difference between the geothermal and control cases. Note that each panel is centered 
on the anti-stellar point (at 0$^{\circ}$ latitude and 0$^{\circ}$ longitude) to emphasize the changes 
in temperature and wind that occur across the anti-stellar hemisphere. A noticeable shift in wind patterns 
occurs between fast and slow rotators, which is an expected regime transition 
\citep{merlis2010,edson2011,yang2014,leconte2013} that 
will be discussed further in Section \ref{section:dynamics}. Asymmetries also persist between the northern 
and southern hemisphere in all cases, which has been noted by others \citep{merlis2010,yang2013,yang2014} 
and also will be discussed further in Section \ref{section:dynamics}. The presence of geothermal heating 
(Fig. \ref{fig:surftemp}, bottom row) provides warming across the surface as expected; however, the 
distribution of warming yields regions near the anti-stellar point with asymmetric  
warm or cold features between the northern and southern hemispheres that persist 
even after averaging over 1000 days. These effects are most pronounced 
in the fast dry cases, with deviations of $\pm$5 K and asymmetric wind patterns around the anti-stellar point. 
The presence of moisture appears to reduce the asymmetric influence of geothermal heating, although some 
asymmetric features between the northern and southern hemispheres still persist. The slow rotators 
show the most uniform effect from geothermal heating, 
with only slight asymmetries in the time average of surface temperature and horizontal wind. 

\begin{figure*}
  \includegraphics[width=40pc]{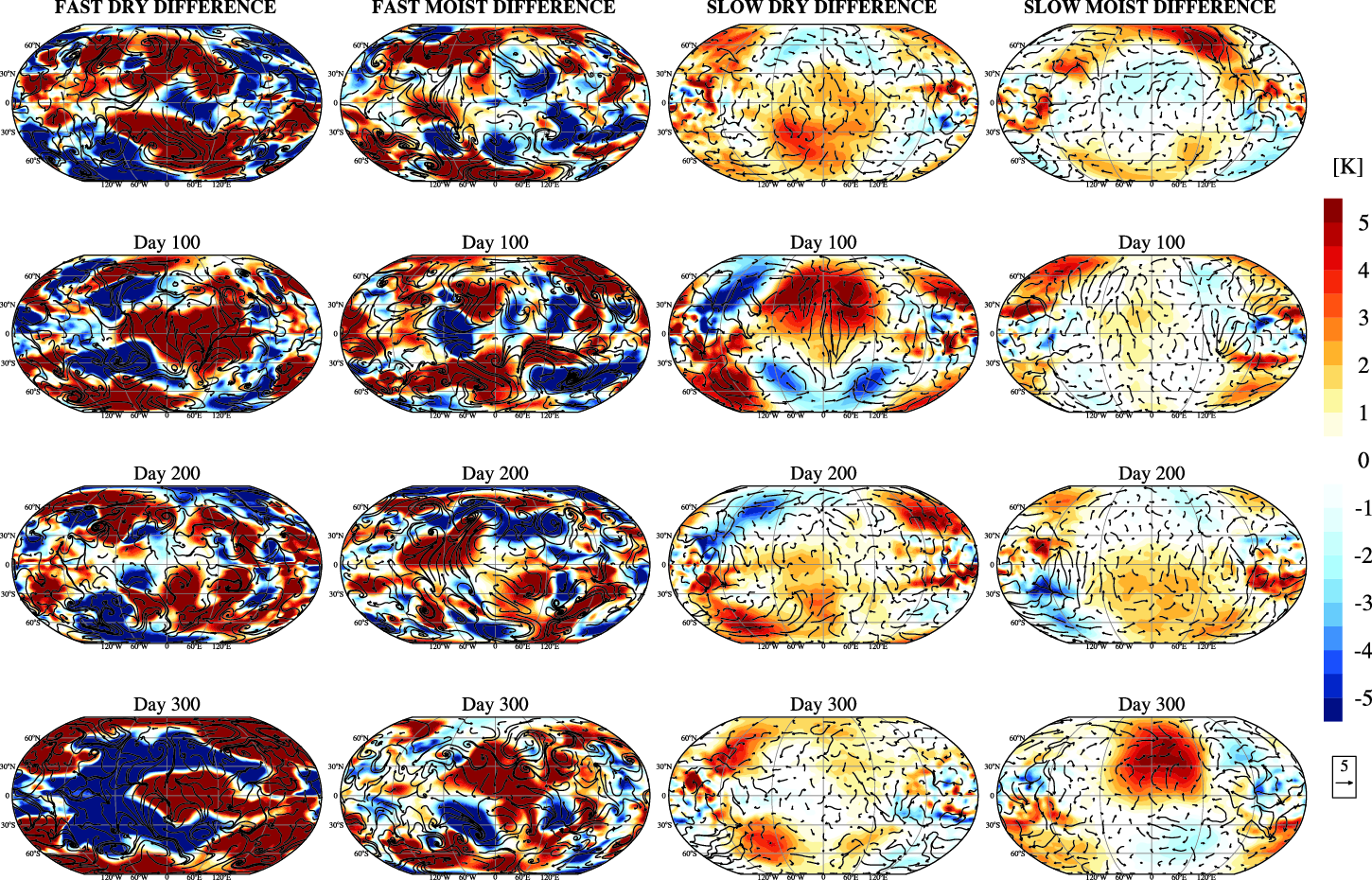}
  \caption{Transient evolution of the surface temperature deviation from the freezing point of water 
(shading) and horizontal wind $u,v$ (vectors) difference between the geothermal and control 
cases for the (first column) fast dry, (second column) fast moist, (third column) slow dry, 
and (fourth column) slow moist experiments. Each row shows instantaneous conditions 
at intervals of 100 days. (Animations of these experiments are available online.)}
  \label{fig:surftempevoall}
\end{figure*}

The asymmetric features between the northern and southern hemispheres 
observed in the time average quantities shown in Fig. \ref{fig:surftemp} are the result  
of transient variations in atmospheric dynamics across the anti-stellar hemisphere. The transient evolution 
of surface temperature and horizontal wind difference (geothermal minus control) 
is shown in Fig. \ref{fig:surftempevoall} at 100-day intervals for the full set of experiments. (Animations 
of transient evolution for all GCM experiments are available online.) In all cases, geothermal heating shows stronger 
regions of warming and cooling that are asymmetrically distributed between the northern and southern hemispheres 
near the anti-stellar point and 
fluctuate with time. The fast dry case shows the strongest deviations of $\pm$20 K or more, while the presence 
of moisture reduces these changes by about a factor of two. Asymmetric features are also prominent in 
the slow dry case, with surface temperature changes of about $\pm$5 K or so and slight reductions in the 
corresponding fast moist case. These experiments all suggest that geothermal heating acts to create or amplify 
asymmetric circulation features between the northern and southern hemispheres that cause 
transient regions of warming and cooling most prominent near the anti-stellar point. 

Geothermal heating enhances the persistence of asymmetric features by affecting the formation 
of large-scale dynamical structures responsible for energy transport. Even in a dry atmosphere, 
strong thermally-direct circulations (discussed further in Section \ref{section:dynamics}) are sensitive to changes 
in the surface energy flux, which alters the transport of mass and energy from the sub-stellar to anti-stellar 
point and generates a slightly different pattern in the general circulation. The net effect is that 
the added energy from uniform geothermal heating is distributed unevenly across the surface through 
dynamical energy transport, which allows some additional cycles of warming and cooling to persist 
between the northern and southern hemispheres near the anti-stellar point.

Even without geothermal heating, asymmetric features between the northern and southern hemispheres 
have been observed in other GCM simulations 
\citep{merlis2010,yang2013,yang2014} and appear to be a robust phenomenon of synchronous 
rotation. We even attempted to initialize the GCM into a fully symmetric synchronously rotating 
state without any baroclinic eddies (following the approach of \citet{haqqmisra2011} 
for non-synchronous rotation), but the model immediately lost its symmetry and reverted 
to the states presented here. These asymmetric features seem to persist regardless of 
the averaging period or sampling variability and do 
not appear sensitive to other changes in the model configuration.
These results all suggest that the asymmetric features seen in 
GCMs are not artifacts of averaging but are dynamically-induced features of the atmosphere that affect 
the distribution of surface temperature between the northern and southern hemispheres, 
particularly near the anti-stellar point.

\section{ATMOSPHERIC CIRCULATION}\label{section:dynamics}

The atmospheric circulation on synchronously rotating planets is known to 
exhibit unique three-dimensional behavior that is markedly different from 
the prominent circulations on Earth, such as equatorial superrotation and heat 
transport from the anti-stellar to sub-stellar point \citep{joshi1997,joshi2003,merlis2010,edson2011,showman2011}. 
Some of this behavior is captured by typical meridional and zonal mass streamfunctions, but 
we show below that careful consideration of the wind field is required to accurately 
describe the most prominent means of transport.

We describe the mean meridional circulation (MMC) with a mass streamfunction 
$\Psi_{M}$ that calculates the northward mass flux above a particular pressure 
level $p^{\prime}$ as
\begin{equation}
\Psi_{M}=\frac{2\pi a\cos\phi}{g}\int_{0}^{p^{\prime}}\bar{v}dp.\label{eq:MMC}
\end{equation}
This streamfunction traces out the familiar patterns of the MMC when
applied to Earth models or time-averaged observations, showing direct
(i.e., Hadley) and indirect (i.e., Ferrel) circulation cells. On Earth, 
and other planets not in synchronous rotation, the MMC is a primary driver of meridional 
heat transport from the tropics to poles and results in the formation of a subtropical jet 
and polar front jet along the edge of the circulation cells \cite{frierson2006}. 

\begin{figure*}
  \includegraphics[width=40pc]{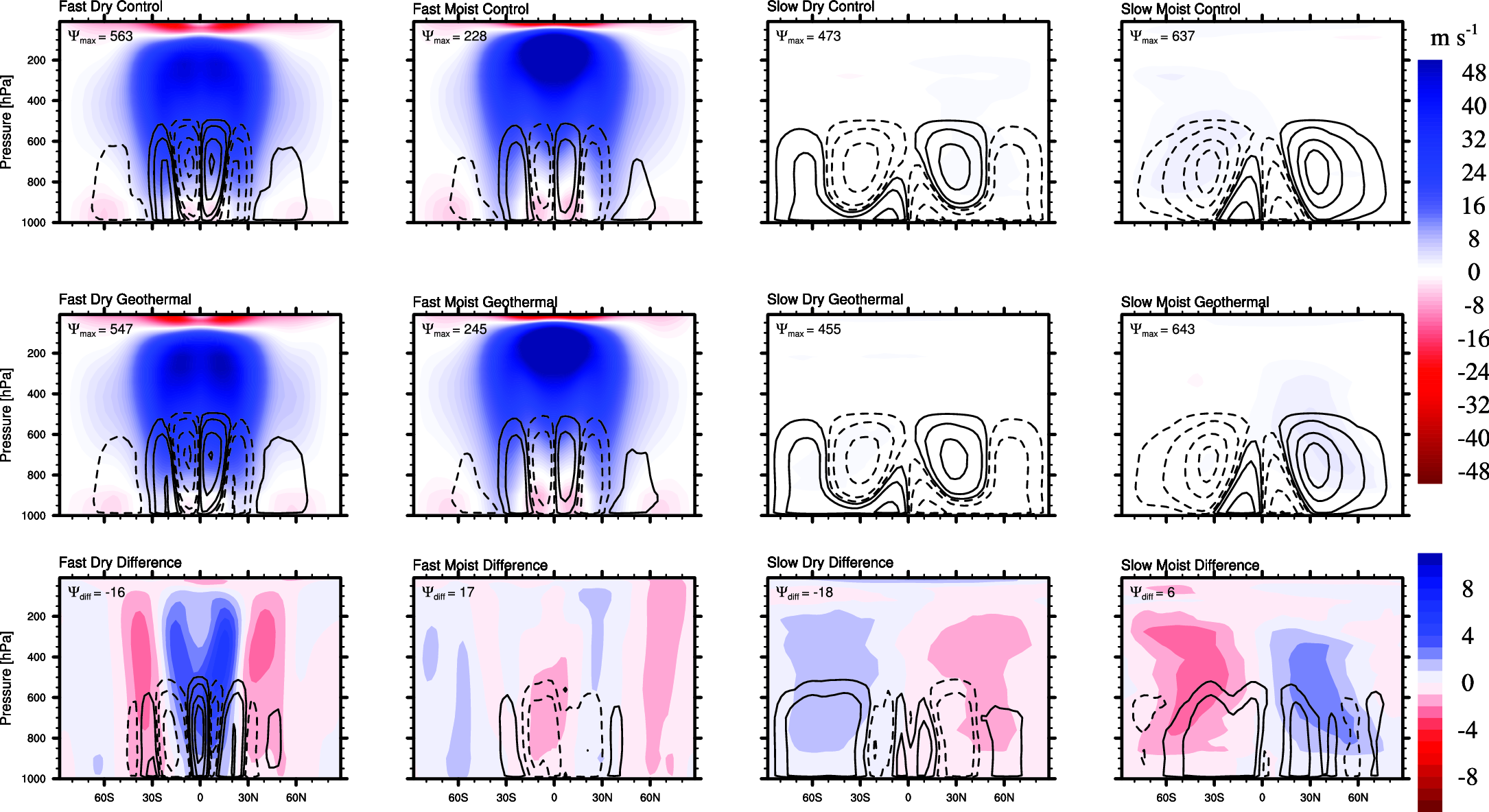}
  \caption{Mean meridional circulation $\Psi_{M}$ (line contours) and zonal
mean zonal wind $\overline{u}$ (shading) for the (first column)
fast dry, (second column) fast moist, (third column) slow dry, and (fourth column) slow moist
experiments. The top row shows control cases, 
the middle row shows geothermal cases, and the bottom row shows the
difference between geothermal and control cases. Contours are drawn 
at an interval of $\pm\{20, 100, 300, 500, 700, 1000\}\times10^{9}$ kg s$^{-1}$ for the 
top two rows and at $\pm\{5, 20, 60, 100\}\times10^{9}$ for the bottom row. 
The maximum streamfunction $\Psi_{max}$ is shown on each panel
in units of $10^{9}$ kg s$^{-1}$. Solid contours indicate clockwise circulation, and dashed contours
indicate counter-clockwise circulation.}
  \label{fig:umeanMMC}
\end{figure*}

For planets in synchronous rotation, the MMC still acts as a means of meridional heat transport, 
but the geometry of the circulation depends strongly upon rotation rate. Fig. \ref{fig:umeanMMC} 
shows the MMC $\Psi_{M}$ and zonal mean zonal wind $\bar{u}$ for the full set of experiments. 
The fast rotators all show similar patterns of a Hadley, Ferrel, and polar cell similar to 
Earth, but the upper atmosphere features strong superrotation in a single eastward equatorial 
jet with westward flow only at the very top of the model. 
The slow rotators show an entirely different MMC configuration, with direct cells elevated 
above the surface and indirect cells circulating from equator to pole. The zonal wind for the 
slow rotators appears to be zero, which we show below results from a change in circulation across 
hemispheres. Note that the difference between geothermal and control cases (Fig. \ref{fig:umeanMMC}, bottom row)
results in asymmetric structures that are weak in strength but emphasize the role of 
geothermal heating in creating small asymmetries in circulation 
between the northern and southern hemispheres. The strength of the MMC 
is also stronger in dry cases compared to their moist counterparts, which occurs 
because the presence of water vapor increases static stability and decreases
the magnitude of overturning circulations \citep{haqqmisra2011}.

The MMC shown in Fig. \ref{fig:umeanMMC} can be misleading, however, as the circulation pattern 
changes depending on position relative to the sub-stellar point. This observation was made 
by \citet{geisler1981} in application to a simplified model of the Walker circulation, where it was noticed 
that the direction of the MMC changes on either side of the heating source. On Earth, the 
Walker circulation is a transient phenomenon, but on synchronously rotating planets the 
star shines fixed on a single location at all times, so this theoretical prediction may be 
even more relevant. We therefore may expect that the patterns 
of circulation in our GCM experiments should change between the hemispheres east and west 
of the sub-stellar point. 

\begin{figure*}
  \includegraphics[width=40pc]{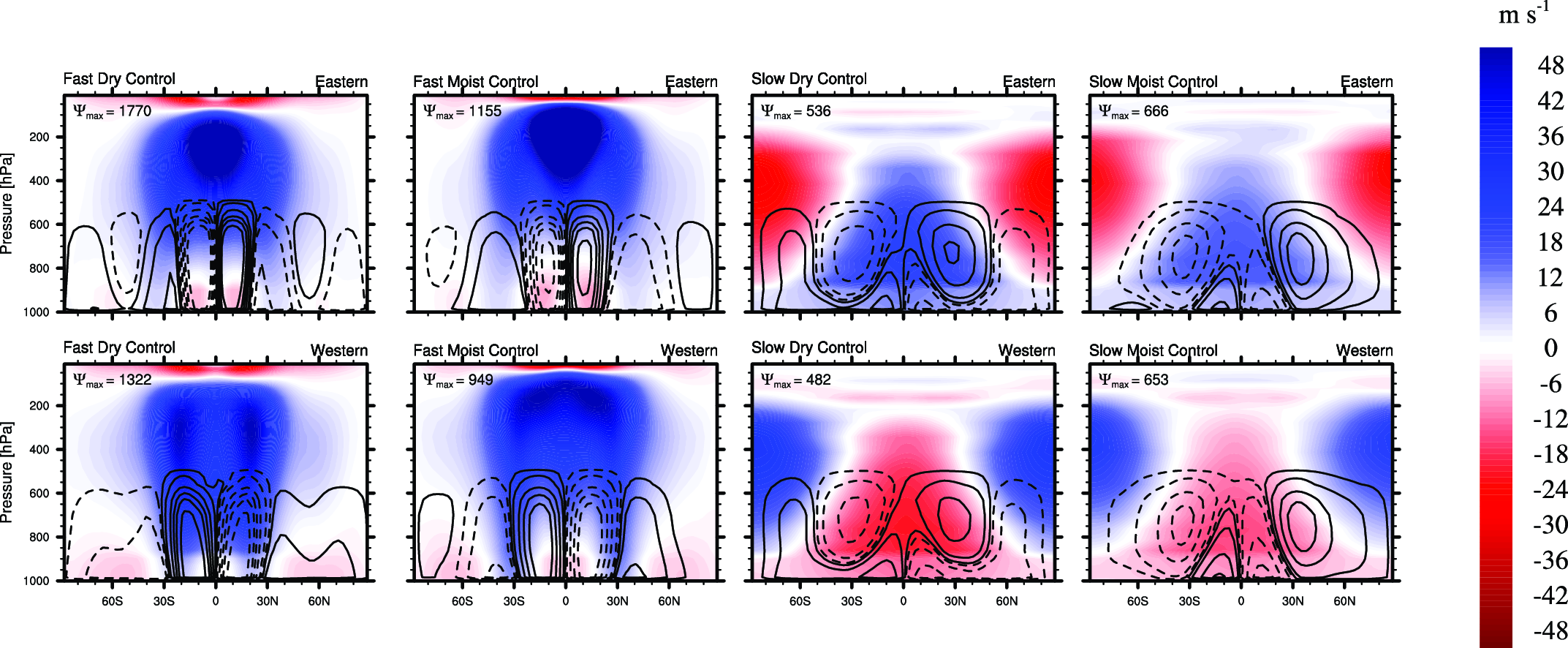}
  \caption{Mean meridional circulation $\Psi_{M}$ (line contours) and zonal
mean zonal wind $\overline{u}$ (shading) averaged across the (top row) eastern hemisphere and (bottom row) western 
hemisphere from the sub-stellar point for the (first column)
fast dry, (second column) fast moist, (third column) slow dry, and (fourth column) slow moist control 
experiments.  Contours are drawn at an interval of $\pm\{20, 100, 300, 500, 700, 1000\}\times10^{9}$ kg s$^{-1}$,
and the maximum streamfunction $\Psi_{max}$ is shown on each panel
in units of $10^{9}$ kg s$^{-1}$. Solid contours indicate clockwise circulation, and dashed contours
indicate counter-clockwise circulation.}
  \label{fig:umeanMMChemi}
\end{figure*}

Fig. \ref{fig:umeanMMChemi} shows the MMC $\Psi_{M}$ and zonal mean zonal wind 
$\bar{u}$ separated over the eastern hemisphere and western hemisphere from the 
sub-stellar point for the set control of experiments. The fast rotators all show a change 
in direction of the MMC, with the eastern hemisphere showing circulation in the same direction as in 
Fig. \ref{fig:umeanMMC} and the western hemisphere showing circulation in the opposite direction. 
The atmosphere is still superrotating for the fast rotators, although the equatorial jet is much 
stronger in the eastern hemisphere, and notable differences persist between hemispheres in the geometry 
of the MMC cells. By contrast, the slow rotators do not show a change in circulation direction, but instead 
the wind direction changes from eastward in the eastern hemisphere to westward in the western 
hemisphere. Slight asymmetries are apparent in the MMC between the northern and southern hemispheres, but the slow rotators 
appear to maintain a relatively consistent geometry across the sub-stellar point. The contrast 
between the zonally averaged quantities in Fig. \ref{fig:umeanMMC} and the hemispheric separation 
made in Fig. \ref{fig:umeanMMChemi} provides one example of the circulation features 
that contribute to persistent atmospheric asymmetries. 

The zonal circulation, also known as the Walker circulation, provides another possible 
mechanism for transporting energy from the sub-stellar to anti-stellar point. On Earth, the 
Walker circulation is characterized by rising motion at the heating source, eastward (and westward) 
flow aloft as air descends, and return flow along the surface back toward the heating 
source \citep{gill1980,geisler1981}. We describe the mean zonal circulation (MZC) with a mass streamfunction 
$\Psi_{Z}$ that calculates the eastward mass flux above a particular pressure 
level $p^{\prime}$ as
\begin{equation}
\Psi_{Z}=\frac{2\pi a}{g}\int_{0}^{p^{\prime}}\bar{u}^{*}dp,\label{eq:MZC}
\end{equation}
where $\bar{u}^{*}$ is the zonal average value minus the time-average component of zonal wind. 
The extent of this circulation is limited by the propagation 
of equatorially trapped Rossby waves and equatorial Kelvin waves \citep{gill1980,showman2011}, 
which provides an explanation for the regime transition in wind patterns observed between 
fast and slow rotators in Fig. \ref{fig:surftemp}. \citet{edson2011} describes a change in 
patterns of winds that occurs when the Rossby deformation radius (which is inversely proportional 
to rotation rate) begins to approach the planetary radius at a rotation rate of about 4.2 days. This 
regime shift has also been noted by others \citep{merlis2010} and is evident in 
many different GCMs \citep{yang2014,leconte2013}. At least for the slow rotators, we expect that 
the zonal circulation should provide a circulation that reaches from the sub-stellar to the 
anti-stellar point. 

\begin{figure*}
  \includegraphics[width=40pc]{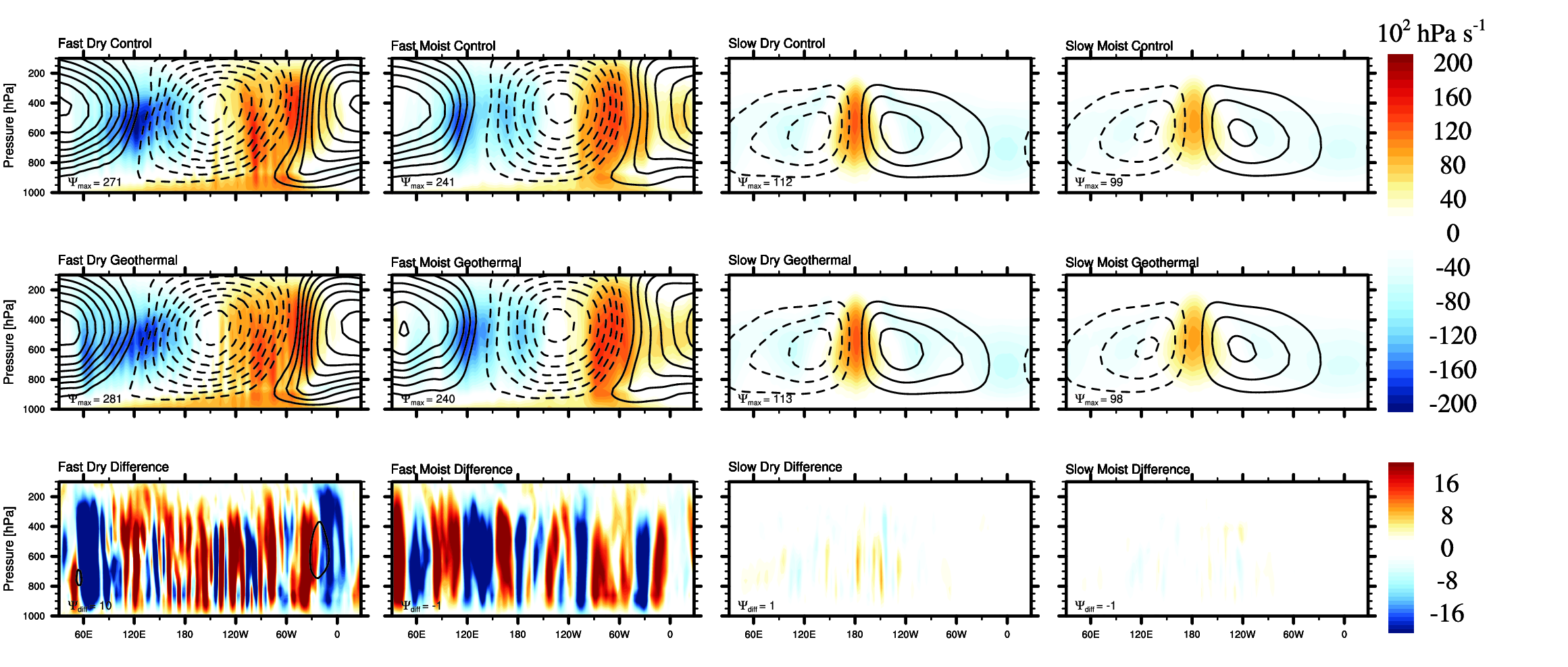}
  \caption{Mean zonal circulation $\Psi_{Z}$ (line contours) and vertical wind
$-\overline{\omega}$ (shading) for the (first column)
fast dry, (second column) fast moist, (third column) slow dry, and (fourth column) slow moist
experiments. The contour interval for the
mean zonal circulation is $30\times10^{11}$ kg s$^{-1}$ for the fast rotators and $30\times10^{13}$ kg s$^{-1}$ for the slow rotators, 
and the maximum streamfunction $\Psi_{max}$ is shown on each panel in units
of $10^{11}$ kg s$^{-1}$ and $10^{13}$ kg s$^{-1}$, respectively. Solid contours indicate clockwise circulation, and dashed contours
indicate counter-clockwise circulation. Zonal wind speeds have been reduced in the slow 
rotator cases by a factor of 10.}
  \label{fig:omegaMZC}
\end{figure*}

Fig. \ref{fig:omegaMZC} shows the MZC $\Psi_{Z}$ and vertical wind $-\bar{\omega}$
for the full set of experiments. The slow rotators do indeed show a characteristic 
Walker-like circulation with rising motion at the sub-stellar point, eastward (and westward) 
motion aloft, sinking motion near the sub-stellar point, and return flow along the surface. 
The fast rotators show the opposite sense: sinking motion in the vicinity near the sub-stellar 
point, and rising motion within the anti-stellar hemisphere. Because the Rossby deformation 
radius is small compared to the planetary radius, the MZC cells for the fast 
rotators span a shorter longitudinal extent and does not fully reach the anti-stellar point. 
Conversely, the slow rotators have a larger Rossby deformation radius that allows the 
MZC to traverse a full hemisphere. Geothermal heating appears to have relatively negligible 
effects on the MZC in all cases, while variations in vertical wind due to geothermal heating 
appear only in the fast rotators.

\begin{figure*}
  \includegraphics[width=40pc]{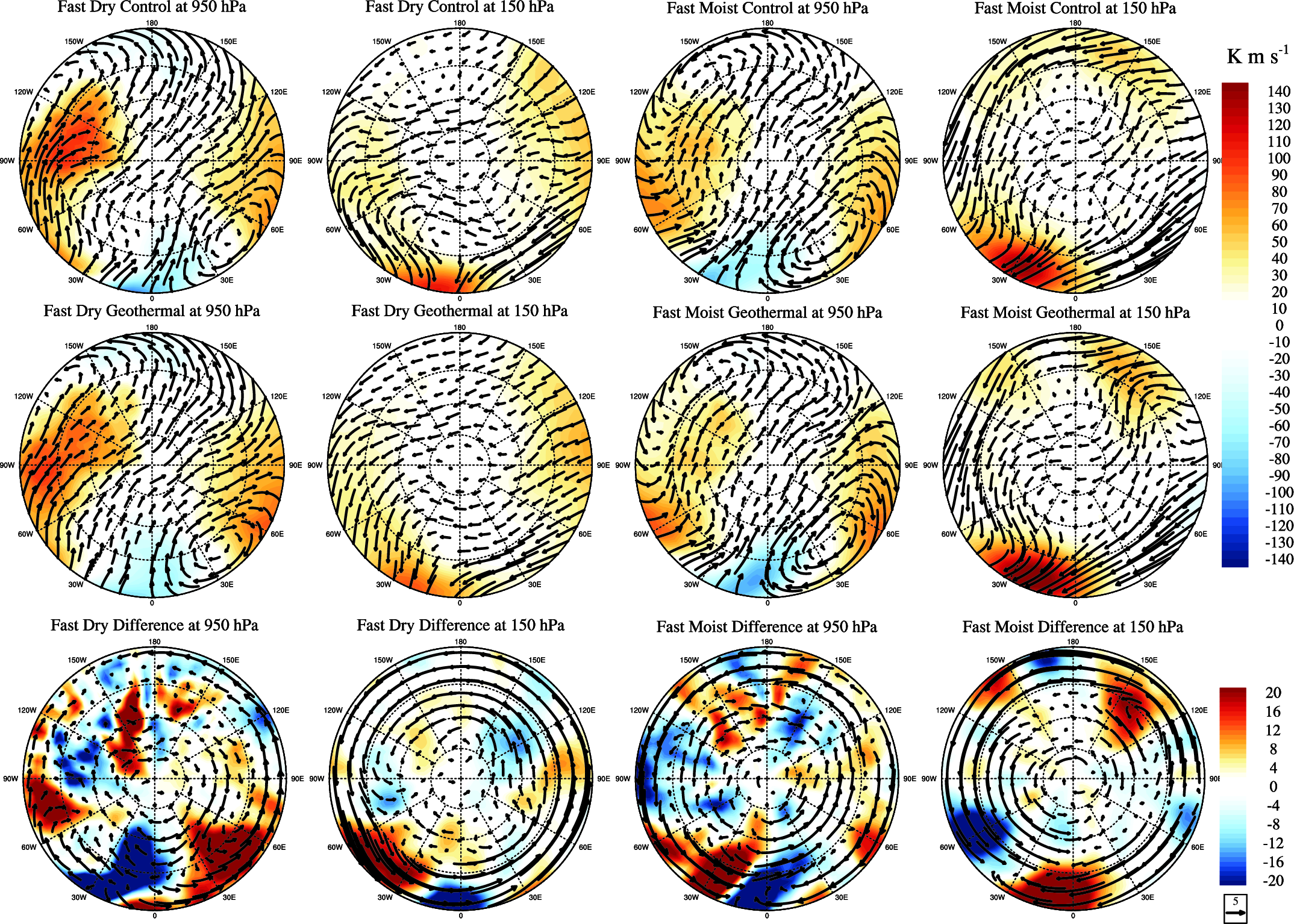}
  \caption{Polar stereographic plots of eddy heat flux $\overline{v'\theta'}$
(shading) and horizontal wind $\overline{u},\overline{v}$ (vectors)
on the (first and third columns) 950 hPa surface and (second and fourth columns) 150 hPa surface
for the (first and second columns) fast dry and (third and fourth columns) fast moist experiments.
The top row shows control cases, the middle row shows geothermal cases, and the bottom row shows the
difference between geothermal and control cases. The plot shows the
northern hemisphere from 30$^{\circ}$ to 90$^{\circ}$ latitude,
and the zonal mean is removed from all zonal wind vectors.}
  \label{fig:polarHeatfast}
\end{figure*}

\begin{figure*}
  \includegraphics[width=40pc]{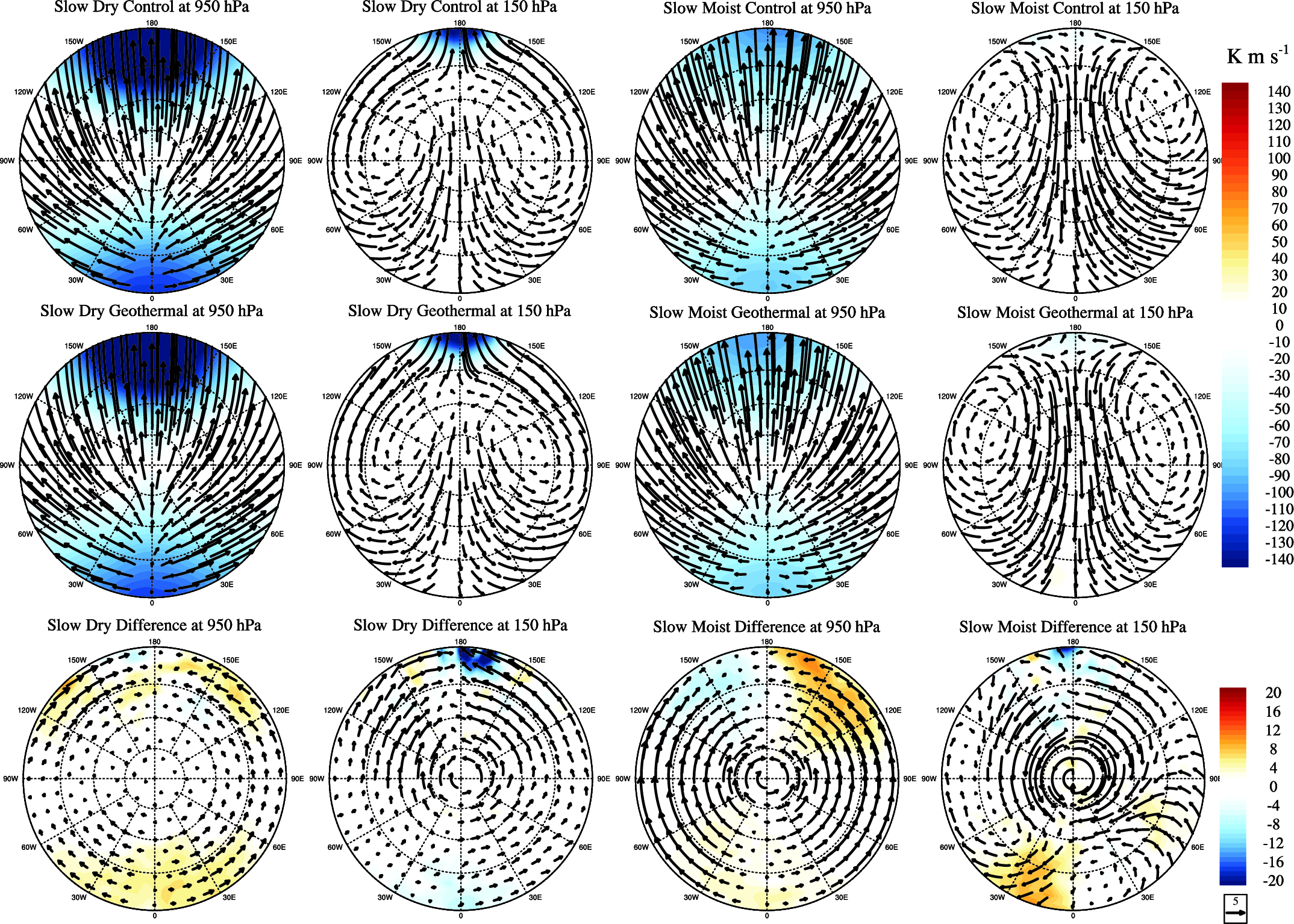}
  \caption{As with Fig. \ref{fig:polarHeatfast}, but for the slow rotator cases.}
  \label{fig:polarHeatslow}
\end{figure*}

The MMC and MZC provide a partial description of the circulation on synchronously rotating 
planets, but the three-dimensional features noted by others \citep{joshi1997,joshi2003,edson2011} 
include a cross-polar circulation that appears in all our GCM experiments and acts as 
an additional means of transport from the sub-stellar to anti-stellar point. 
Fig. \ref{fig:polarHeatfast} shows a polar stereographic plot of eddy heat flux 
$\overline{v'\theta'}$ and horizontal wind for the set of fast 
rotators, where $\theta$ is potential temperature and the primes denote deviations 
from the zonal mean. Following \citet{joshi1997}, the zonal mean also has been removed 
from all zonal wind vectors. The first and third columns of Fig. \ref{fig:polarHeatfast} show 
the 950 hPa surface just above the ground, while the second and fourth columns 
show the 150 hPa surface at the top of the troposphere. In both dry and moist cases, 
a cross-polar circulation is evident with flow from the anti-stellar point to sub-stellar 
point along the surface, rising motion at the heating source, and flow aloft from the 
sub-stellar point back across the pole to the anti-stellar point. Likewise, 
Fig. \ref{fig:polarHeatslow} shows a polar stereographic plot of eddy heat flux 
and horizontal wind for the set of slow rotators, which shows the same cross-polar circulation 
pattern with more pronounced patterns of superrotation. Geothermal heating helps to 
increase the magnitude of superrotation in all these experiments, which increases 
the transport of mass and energy to the anti-stellar point by enhancing the cross-polar flow.

\begin{figure}
  \includegraphics[width=20pc]{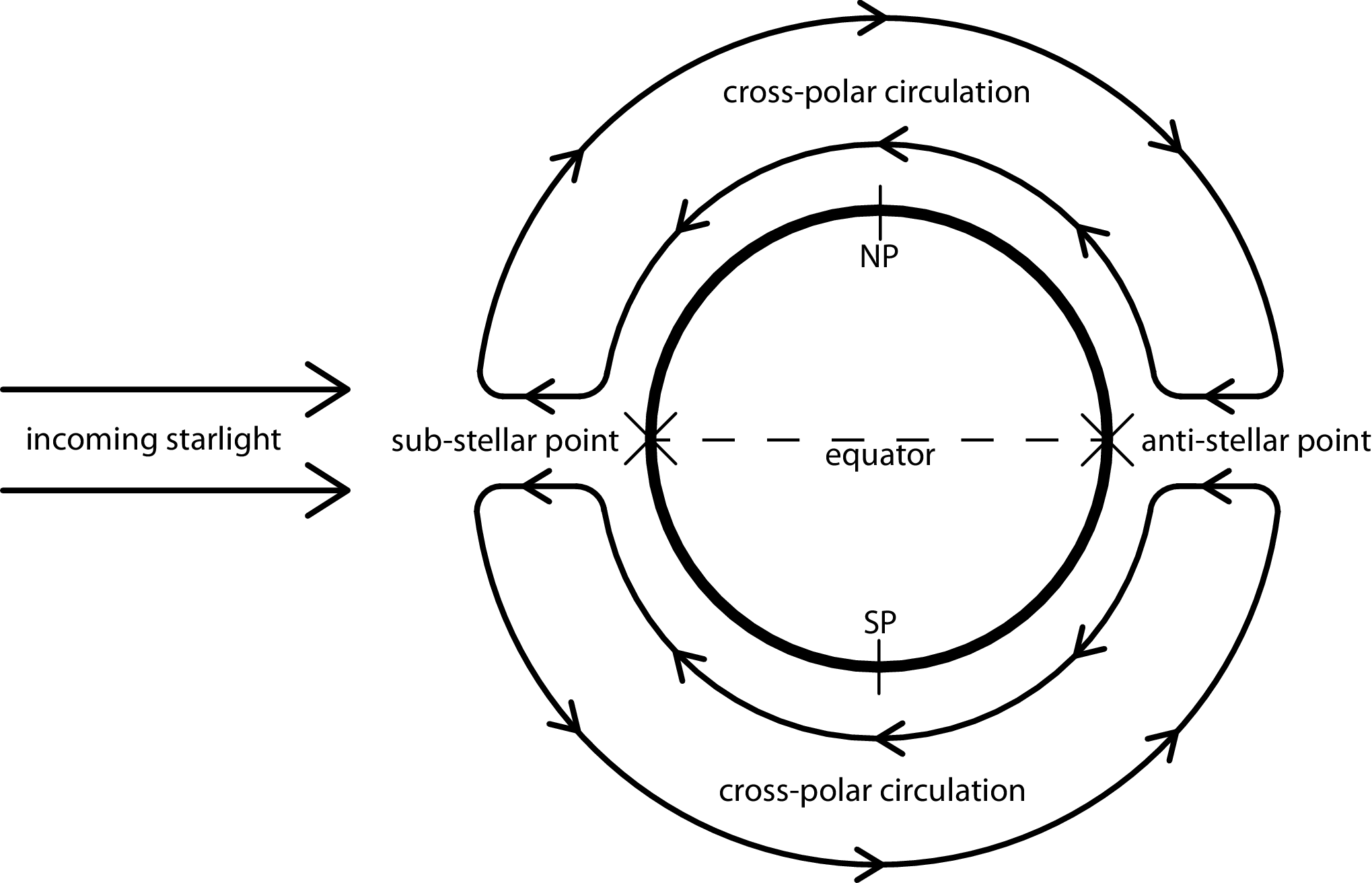}
  \caption{A schematic diagram of the cross-polar circulation that provides an additional means of  
transport from the sub-stellar to anti-stellar points on synchronously rotating planets.}
  \label{fig:geofluxcartoon}
\end{figure}

The cross-polar circulation appears to be evident as a means of transport from the sub-stellar 
to anti-stellar point in all these experiments. The MMC and MZC also provide prominent circulations 
that maintain transient asymmetries between the northern and southern hemispheres,
although only in the slow rotators where the Rossby 
deformation radius is large does the MZC also contribute to this transport. 
Fig. \ref{fig:geofluxcartoon} shows a schematic diagram of the cross-polar circulation 
that contributes to energy transport on synchronously rotating planets.  This polar flow is present 
in all our experiments, regardless of rotation rate or geothermal heating, and should be a robust 
circulation feature in the atmospheres of synchronously rotating Earth-like planets.

\section{IMPLICATIONS FOR HABITABILITY}

GCM studies have shown that synchronously rotating planets should be stable against 
atmospheric collapse as long as the atmosphere is sufficiently thick 
\citep{joshi1997,joshi2003,merlis2010,edson2011}. The dry and moist GCM 
calculations in this study all specify an Earth-like value of reference surface pressure 
(1000 hPa), which should be resilient to collapse for any value of rotation rate (or 
at least until the planet becomes close enough to its host star to trigger 
a runaway greenhouse \citep{barnes2013}). The interpretation of these results therefore 
should be robust across any intermediate rotation rates between the fast and slow 
cases considered here, and similar patterns of large-scale dynamics should be evident 
in other GCMs as well.

The presence of a geothermal heat flux due to tidal heating has the greatest impact 
on the fast rotators. The cross-polar circulation provides enhanced warming on the 
anti-stellar side of the planet, which creates transient regions of warming that can 
reach or exceed 20 K at times. This may be sufficient warming to initiate ice melt and provide 
warmer conditions on a transient basis, and in extreme cases this could even result in the 
permanent presence of standing liquid water on the anti-stellar side of the planet. 
The degree to which this enhanced heat flux could melt ice will depend on the physical properties 
of each particular planet, and representing this complex ice flow is beyond the scope of 
this study. Nevertheless, the presence of a geothermal heat flux on fast rotators may 
contribute to preventing hard glaciation of the anti-stellar hemisphere and increase 
the available habitable surface area.

Slow rotators should have greatest habitability potential along the terminator. 
Transient effects due to geothermal heating provide only a few degrees of warming 
or cooling, but this still might be sufficient to cause patterns in ice 
that forms on the anti-stellar side of such planets. Whether or not this will result 
in permanent ice melt will depend on physical properties of the planet, but the 
strong superrotation on slow rotators tends to dominate the large-scale 
dynamical patterns. The cross-polar circulation on slow rotators complements 
the zonal circulation so that surface winds everywhere are drawn into the 
sub-stellar point and carried aloft toward the anti-stellar point. 

In general, the presence of the cross-polar circulation provides an additional 
mechanism for transporting energy and mass from the sub-stellar to anti-stellar 
point. Linearized theory based on the analysis of the Walker circulation on 
Earth suggests analogous structures for the circulation on synchronous 
rotators \citep{gill1980,geisler1981,showman2011}, but these theoretical expectations
of a zonal circulation reaching from the sub-stellar to anti-stellar point only 
seem to apply to slow rotators where the Rossby deformation radius is large. 
Faster rotators, with a smaller Rossby deformation radius, do not exhibit 
significant zonal transport, but the pattern of the cross-polar circulation perhaps 
is more consistent with theory. This suggests that any GCM simulations 
(and, someday, observations) of synchronously rotating planets should consider 
the mass transport by the cross-polar circulation, rather than the meridional or zonal circulations exclusively, 
as a means of characterizing the dynamics of climate. 

The climate of a synchronously rotating planet is sometimes described using the term
``eyeball Earth'' to denote open water beneath the sub-stellar point, relatively clement 
conditions along the surface in a surrounding ring, and ice-covered conditions everywhere 
else all the way to the anti-stellar hemisphere \citep{pierrehumbert2011,angerhausen2013}. 
If geothermal heating can cause regions of transient warming 
through the enhancement of large-scale circulations, then regions of melting could create cratered pockets 
of ice, slowly flowing glaciers, or even standing liquid water around the anti-stellar point. 
This suggests that ``eyeball Earth'' planets might include a soft spot on the side farthest 
from the heat source. Further investigation with GCM's coupled to ice sheet models will 
help to constrain the range of climates on which this could occur.

\section{CONCLUSIONS}

Low mass M-type stars provide some of the closest and most numerous targets 
for current exoplanet surveys, and the discovery of synchronously rotating planets 
within the habitable zones of these stars will increase interest in the 
atmospheric characterization of such worlds. We have considered a set of experiments 
that include slow rotators (1 day) toward the inner edge of the habitable zone and 
fast rotators (230 days) toward the outer edge of the habitable zone. 
Between these fast and slow rotators is a regime transition where superrotation dominates, 
but the general patterns discussed from this limited set of experiments should extend 
appropriately to other rotation rates within the habitable zone.

Tidal heating can produce a geothermal heat flux that warms the surface of a synchronously 
rotating planet, which can cause transient regions of warming and cooling of up to $\pm$20 K or 
more on the anti-stellar hemisphere. This enhanced warming could contribute to increased ice 
melt along the anti-stellar point and extend the habitable surface area of the planet. This 
effect is stronger on fast rotators, and the presence of moisture also tends to reduce the 
magnitude of anomalous regions. The presence of geothermal heating acts to enhance 
asymmetries between the northern and southern hemispheres that alter the structure 
of large-scale circulation features such as the MMC, 
the MZC, and the cross-polar circulation.

The dynamics of synchronous rotators include different patterns on either side of the sub-stellar point. 
The fast rotators exhibit a change in the direction of the Hadley circulation from the eastern to 
western hemisphere, while the slow rotators show a change in the direction of zonal wind. The Walker 
circulation provides transport of mass and energy from the sub-stellar to anti-stellar point only for slow rotators, 
while fast rotators show a Walker circulation with the opposite sense. In all cases, a 
cross-polar circulation provides an additional mechanism for transport of mass and energy 
from the sub-stellar to anti-stellar point. Fast rotators maintain asymmetric features such as 
large-scale jets and vorticies at midlatitudes, while slow rotators have transitioned to 
a superrotation regime, but all experiments show a cross-polar circulation that provides 
or enhances the transport from day side to night side. Typical considerations of the Hadley and 
Walker circulation therefore may be insufficient for synchronously rotating planets, with the 
cross-polar circulation and hemispheric meridional circulation providing more accurate descriptions 
of atmospheric transport.

Our set of experiments is designed to highlight important features in the large-scale dynamics 
of synchronously rotating planets. The idealized GCM used in this study cannot provide precise  
orbital limits to habitability, but the simplified nature of this model allows the most fundamental 
atmospheric processes to be more apparent. More accurate representation of clouds and radiative 
transfer will improve the model's ability to predict particular climates for specific planetary 
configurations, but the general patterns of large-scale dynamics should remain consistent 
across the parameter space described here. As astronomers continue the hunt for alien 
worlds and turn their eyes to low mass stars nearby, atmospheric modelers should continue 
the quest to understand circulation patterns on worlds very much unlike our own.

\section*{Acknowledgments}
Funding for this research was provided by the NASA Astrobiology Institute's Virtual
Planetary Laboratory under awards NNX11AC95G,S03 (J.H.) and NNH05ZDA001C (R.K.K.).

\label{lastpage}

\end{document}